\documentclass[prd,a4paper,11pt,twoside,superscriptaddress,showpacs,showkeys]{revtex4}
\usepackage[dvipdfm]{graphicx}
\usepackage{array,tabularx}
\usepackage{amsmath}    
\usepackage{verbatim}   
\usepackage[dvipdfm]{hyperref}   

\def\zh2qqh{$ZH \to q\var{q}H$}
\def\zh2qqcc{$ZH \to q\bar{q}c\bar{c}$}
\def\qqqq{q\bar{q}q\bar{q}}
\def\qq{q\bar{q}}

\def\llll{{\ell}{\ell}{\ell}{\ell}}
\def\nlqq{{\nu}{\ell}{q}{q}}
\def\nnqq{{\nu}{\bar{\nu}}{q}{\bar{q}}}
\def\llqq{{\ell}{\ell}{q}{\bar{q}}}

\def\tt{t\bar{t}}

\begin{document}

\title{Evaluation of measurement accuracies of the Higgs boson branching fractions in the International Linear Collider}

\author{H. Ono}
\email{ono@ngt.ndu.ac.jp}
\affiliation{Nippon Dental University School of Life Dentistry at Niigata}
\author{A. Miyamoto}
\affiliation{High Energy Accelerator Research Organization}

\date{\today}


%
\begin{abstract}
 Precise measurement of Higgs boson couplings is an important task
 for International Linear Collider (ILC) experiments
 and will facilitate the understanding of the particle mass generation mechanism.
 In this study,
 the measurement accuracies of the Higgs boson branching fractions to the $b$ and $c$ quarks and gluons,
 $\Delta Br(H\to b\bar{b},~c\bar{c},~gg)/Br$,
 were evaluated with the full International Large Detector model (\texttt{ILD\_00}) for
 the Higgs mass of 120~GeV at the center-of-mass (CM) energies of 250 and 350 GeV using
 neutrino, hadronic and leptonic channels
 and assuming an integrated luminosity of $250~{\rm fb^{-1}}$,
 and an electron (positron) beam polarization of $-80\%$ ($+30\%$).
 We obtained the following measurement accuracies of the
 Higgs cross section times branching fraction ($\Delta (\sigma \cdot Br)/\sigma \cdot Br$)
 for decay of the Higgs into $b\bar{b}$, $c\bar{c}$, and $gg$; as
 1.0\%, 6.9\%, and 8.5\% at a CM energy of 250 GeV and
 1.0\%, 6.2\%, and 7.3\% at 350 GeV, respectively. 
 After the measurement accuracy of the cross section ($\Delta\sigma/\sigma$)
 was corrected using the results of studies
 at 250 GeV and their extrapolation to 350 GeV,
 the derived measurement accuracies of the branching fractions ($\Delta Br/Br$)
 to $b\bar{b}$, $c\bar{c}$, and gg were
 2.7\%, 7.3\%, and 8.9\% at a CM energy of 250 GeV and
 3.6\%, 7.2\%, and 8.1\% at 350 GeV, respectively. 
\end{abstract}

\maketitle

 \section{Introduction}
  Precise measurement of the Higgs boson branching ratios (BRs) is an important task for the International Linear Collider (ILC) program.
  It is also crucial for the understanding of the nature of electro-weak symmetry breaking and
  provides a window to investigate physics beyond the standard model (SM).
  The relatively low background and well-defined initial state of the ILC experiments allow precise, model-independent study of the Higgs boson, which is not an easy task for Large Hadron Collider experiments~\cite{ATLAS, CMS}.
  Measurements of the Higgs BRs to $b\bar{b}$ and
  $c\bar{c}$ decays at an $e^{+}e^{-}$ linear collider were reported in Refs.~\cite{RDR, Higgs1, Higgs2, Higgs3, SiD_qqH}.
  In this study, we investigate the accuracies of BRs of the Higgs to
  $b\bar{b}$, $c\bar{c}$, and $gg$ using \texttt{Geant4}~\cite{Geant4} based realistic simulation implemented with a proposed International Large Detector (ILD)~\cite{ILD}.

  In this study, we assume a Higgs mass of 120~$\rm GeV/c^{2}$ and an integrated luminosity of 250~$\rm fb^{-1}$, and estimate the
  accuracies of the BRs at center-of-mass (CM) energies of 250 and 350 GeV.
  The former value is close to the threshold of Higgs production and thus is considered as initial target of ILC experiments.
  The latter is close to the threshold of top quark pair production; therefore, Higgs data can be corrected
  simultaneously with a top threshold study.
  The difference between kinematical conditions at 250 and 350 GeV could yield different detection efficiencies and
  thus different BR accuracies.
  The accuracies at 250 and 350 GeV under the same conditions are studied and compared.

  The experimental conditions for this study are described in section \ref{sec:ILC}.
  We selected the Higgs events in three channels: neutrino, hadronic and leptonic.
  The event selection and background suppression processes are described in the section \ref{sec:analysis}.
  The derivation of the BRs is presented in the section \ref{sec:template},
  and the conclusion is given in the last section.
  
  \section{Higgs physics in the ILC experiment} \label{sec:ILC}
  
  \subsection{ILC experiment and Higgs production}
  
\begin{figure}[htbp]
  \begin{center}
   \includegraphics[width=0.8\textwidth]{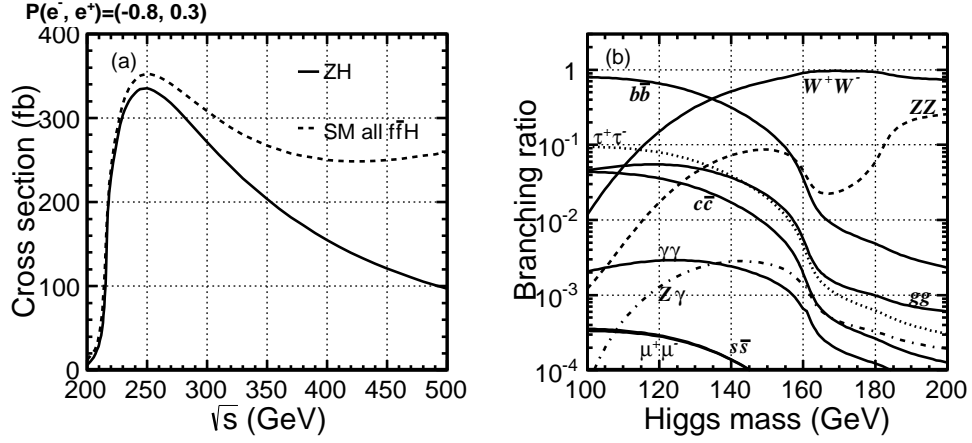}
   \caption{(a) Production cross section of Higgs boson at a Higgs mass of 120 GeV through the Higgs-strahlung ($ZH$) (solid) and all $f\bar{f}H$ (dashed) processes assuming the -80\% electron and
   +30\% positron beam polarization. The cross section is calculated including the initial state radiation by \texttt{Whizard}~\cite{WHIZARD}.
   (b) SM Higgs branching fractions as a function of Higgs mass with \texttt{PYTHIA}~\cite{PYTHIA}.}
    \label{fig:Higgs_xsec_br}
  \end{center}
 \end{figure}

  The ILC is a future electron-positron ($e^{-}e^{+}$) linear collider
  for experiments at an initial center-of-mass (CM) energy ($\sqrt{s}$) up to 500 GeV, which can be extended to 1 TeV.
  The production cross section of the Higgs boson is
  shown in Fig.~\ref{fig:Higgs_xsec_br}(a) as a function of the CM energy for a Higgs mass of 120 GeV.
  At a low CM energy, the Higgs boson is produced primarily through the Higgs-strahlung $e^{+}e^{-} \to ZH$ process,
  which has a maximum around 250 GeV when the effect of the initial state radiation is considered.
  This is about 20 GeV higher than that without the initial state radiation.
  At the $\sqrt{s}=350~{\rm GeV}$, the total cross section is reduced,
  although the contribution of W/Z fusion is greater than that at 250 GeV.
  The decay BRs of the Higgs boson in the SM are shown
  as a function of its mass in Fig.~\ref{fig:Higgs_xsec_br}(b).
  The Higgs decays mainly to $b\bar{b}$ if its mass is below 140~GeV and to $WW^{*}$ in the case of a mass of above 140~GeV.
  
\begin{figure}[htbp]
  \begin{center}
  \includegraphics[width=\textwidth]{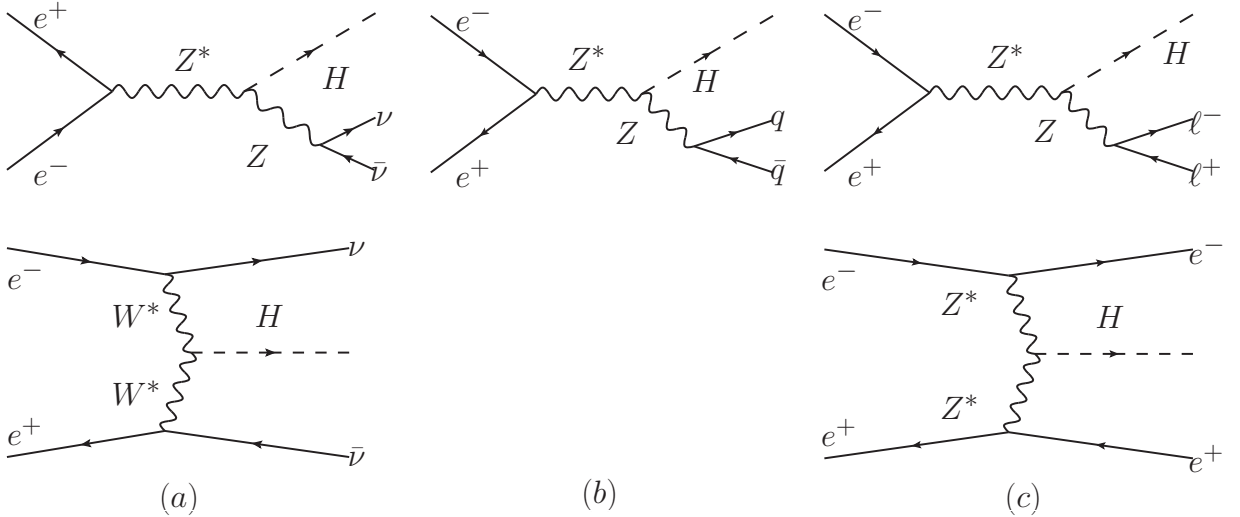}
   \caption{Higgs boson production diagrams categorized according to the final states: (a) neutrino ($\nu\bar{\nu}H$), (b) hadronic ($q\bar{q}H$), and (c) leptonic ($\ell^{+}\ell^{-}H$) channels. Each
   channel is produced mainly through the Higgs-strahlung ($ZH$) process at low CM energies, although the neutrino and leptonic channels also include the $WW$ and $ZZ$ fusion processes, respectively.}
   \label{fig:Higgs_diagram}
  \end{center}
 \end{figure}

 Higgs analysis modes are categorized in terms of the three $Z$ boson decay channels: $Z\to\nu\bar{\nu}$ (neutrino), $q\bar{q}$ (hadronic), and $\ell^{+}\ell^{-}$ (leptonic), as shown in Fig.~\ref{fig:Higgs_diagram}.
 We assumed the $-80\%$ and $+30\%$ polarization of the initial electrons and positrons, respectively, in order to enhance the Higgs signals.
 
 \subsection{ILD concept}
 
\begin{figure}[htbp]
  \includegraphics[width=0.5\textwidth]{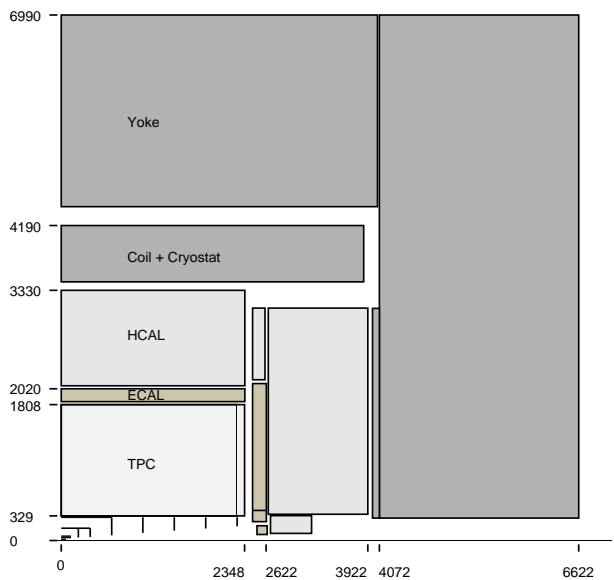}
  \caption{Schematic view of the ILD detector.}
  \label{fig:ILD detector}
 \end{figure}

 We used the ILD~\cite{ILD} model for this study.
 The ILD, which is the validated detector concept for the ILC,
 is equipped with a highly segmented calorimeter and a hybrid tracking
 system consisting of gaseous, silicon-strip, and silicon-pixel trackers. They provide 
 an excellent jet energy resolution by particle flow analysis, as well as excellent momentum resolution and vertex flavor tagging capability, which are necessary  for measuring multi-jet final states in the ILC energy region.
 All sub-detector components of the ILD are shown in Fig.~\ref{fig:ILD detector};
 which consists of silicon-pixel vertex detectors (VTX), silicon inner and outer detectors (SIT, SET), 
 a time projection chamber (TPC),
 high-granularity electromagnetic and hadron calorimeters (ECAL, HCAL),
 a super-conducting solenoid magnet with a $3.5~\rm{T}$ magnetic field,
 and an iron return yoke with a muon detector.
 In addition, forward silicon trackers (FTD, ETD) and beam/luminosity calorimeters (LCAL, LHCAL and BCAL)
 are installed in the forward region.

 The VTX system consists of three double layers of silicon pixel sensors 
 with a 2.8 $\mu {\rm m}$ point resolution located at radii between 16 mm and 60 mm,
 the total radiation length being 0.74\%.
 The impact parameter resolution ($\sigma_{IP}$) of the VTX system
 is $5~{\rm \mu m} \oplus 10~{\rm \mu m \cdot GeV/c}/p \sin^{3/2}{\theta}$.
 The TPC occupies a volume up to a radius of 1.8 m and a half-length in Z of 2.3 m,
 providing a stand-alone momentum resolution of $\sigma_{1/P_{T}} \sim 9 \times 10^{-5}~{\rm GeV^{-1}}$.
 The SIT and SET are placed at the inner and outer sides of the TPC
 with 7 and 50 $\mu \rm m$ point resolutions in the $R-\phi$ and $z$ directions, respectively.
 The overall momentum resolution of the tracking system ($\sigma_{1/P_{T}}$)
 is $2\times 10^{-5}~{\rm GeV^{-1}} \oplus 1\times 10^{-3}/P_{T}\sin{\theta}$
 for the momentum range 1-200 GeV~\cite{ILD}.
 The ECAL consists of 24 $X_{0}$ tungsten absorbers with highly segmented ($\rm 5\times 5~mm^{2}$) readouts.
 The HCAL consists of 5.5 $\lambda_{I}$ steel absorbers with a $\rm 3\times 3~mm^{2}$ scintillator tile readout.
 With the ILD particle flow algorithm package, \texttt{PandoraPFA}~\cite{PandoraPFA}, a dijet energy resolution of  $25\%/\sqrt{E~(\rm GeV)}$ has been achieved for a 45-GeV dijet,
 which corresponds to a single-jet energy resolution of $\sigma_{E_{j}}/E_{j}=3.7\%$~\cite{ILD}.  

 \subsection{Analysis framework and Monte Carlo samples}

  Monte Carlo (MC) generator samples for the physics study were produced using the \texttt{Whizard}~\cite{WHIZARD},
  and fragmentation and hadronization processes were simulated by \texttt{PYTHIA}~\cite{PYTHIA}.
  The SM Higgs branching fractions in \texttt{PYTHIA} are
  65.7\%, 3.6\%, and 5.5\% for $b\bar{b}$, $c\bar{c}$, and $gg$, respectively.
  The generated particles were passed through the \texttt{Geant4}~\cite{Geant4} based detector simulator \texttt{Mokka}~\cite{Mokka}
  with the ILD model.
  The simulated hits were digitized and then reconstructed by the \texttt{MarlinReco} package;
  then, the resulting skimmed data were analyzed.
  The statistics of the simulated Higgs signal samples were 500 $\rm fb^{-1}$ for both CM energies of both 250 and 350 GeV,
  whereas those for background processes varied with the signal-to-noise ratio (S/N).
  They are scaled in the analysis in order to obtain results corresponding to an integrated luminosity of 250 $\rm fb^{-1}$.
  The major SM background processes for the $e^{+}e^{-}\to ZH$ analysis are $e^{+}e^{-}\to ZZ$ and $W^{+}W^{-}$;
  thus we considered final sample states of $\nnqq$, $\nlqq$, $\llqq$, $\nu\nu\ell\ell$, $\qqqq$ and $\llll$.
  In addition, the $\qq$ and $\tt$ backgrounds were also considered for the neutrino and hadronic channels (but only for $\sqrt{s}=350~{\rm GeV}$ because we used a top mass of $\rm 174.9~{GeV/c^{2}}$).
  In the leptonic channel, most of the multi-jet backgrounds are well suppressed if dilepton identification is required;
  thus, only the $\llqq$ and $\nlqq$ backgrounds were considered.
  We used the 250-GeV samples produced for the ILD letter of intent (LOI) studies~\cite{ILD};
  thus, their beam parameters correspond to those defined in the  ILC Reference Design Report~\cite{RDR}.
  On the other hand, the 350-GeV samples were newly produced for this study using the updated beam parameter SB2009~\cite{SB2009}.
  The instantaneous luminosities were 0.75 and 1 $\times (\rm 10^{34}~cm^{-2}s^{-1})$ for 250 and 350 GeV,
  which yield integrated luminosities of 188 and 250 $\rm fb^{-1}$, respectively, for about 3 years at 100 days of operation per year.
  
 \section{Event reconstruction and background suppression} \label{sec:analysis}

 Depending on the $Z$ decay mode,
 the analysis channels are categorized as the
 neutrino (dijet), hadronic (four-jets) and leptonic channels (dileptons + dijets),
 which are described in the following subsections.
 
  \subsection{Neutrino channel ($\nu\bar{\nu}H$)}

  For neutrino channel analysis,
  particles in the event are first forcibly clustered into two jets by the Durham jet-finding algorithm.
  After the dijet clustering, background reductions are applied according to the selection criteria in Table~\ref{table:nnh}.
  At a CM energy of 250 GeV,
  the Higgs is produced almost at rest because it is close to the production threshold,
  whereas it is boosted at 350 GeV.
  Thus, the cut conditions are optimized to obtain the best S/N at each energy.
  In this channel, $Z$ boson decays invisibly ($\nu\bar{\nu}$);
  thus, the $\nnqq$ and $\nlqq$ processes in the SM are the main backgrounds.
  To reduce them,
  a cut on the missing mass ($M_{miss}$) is applied.
  Although this cut decreases the Higgs signal from the WW fusion process,
  the $\nlqq$, $\llqq$ and $\qqqq$ backgrounds are effectively reduced.
  $\qq$ background is reduced by the following kinematical cuts:
  the transverse momentum ($P_{t}$), longitudinal momentum ($P_{l}$), and maximum momentum ($P_{max}$).
  The $\llll$ background is well reduced by a cut on the number of charged tracks in an event ($N_{chd}$).
  In addition, the $\nlqq$ background reduction is improved by the $Y_{12}$ and $Y_{23}$ cuts.
  $Y_{12}$ and $Y_{23}$ are the maximum and the minimum of $y$ values (scaled jet masses), respectively, required to cluster the event into two jets.

  The background reductions for each cut are summarized in Table~\ref{table:nnh} for each CM energy.
  After all selection criteria are met,
  an additional likelihood ratio ($LR$) cut is applied to improve the background reduction.
  The $LR$ is defined using the following variables:
  $M_{miss}$, the number of particles ($N_{PFO}$), $Y_{12}$, $P_{max}$, $P_{\ell}$, and $M_{jj}$.
  The likelihood cut positions are optimized to maximize signal significance
  and $LR > 0.165$ and $LR > 0.395$ are selected for CM energies of 250 and 350 GeV, respectively.
  The signal significance ($S/\sqrt{S+B}$) after all background reductions is also listed in Table~{\table:nnh} with its efficiency,
  where $S$ and $B$ are the  numbers of Higgs signal and background entries, respectively,  after all cuts are applied.
  The remaining backgrounds are $\nlqq$ (60\%), $\nnqq$ (20\%), and $\qq$ (10\%) at both 250 and 350 GeV.
 \begin{table}[htbp]
  \begin{center}
   \caption{Summary of the $\nu\bar{\nu}H$ channel background reduction assuming ${\cal L}=250~{\rm fb^{-1}}$ with $P(e^{-}, e^{+})=(-0.8, +0.3)$.}
   \label{table:nnh}
   \begin{tabular}{|c|c|r|r|c|r|r|}
    \hline
    CM energy (GeV) &\multicolumn{3}{c|}{250} & \multicolumn{3}{c|}{350} \\
    \hline
    Cut names & condition & Sig.& Bkg. & condition & Sig. & Bkg.\\
    \hline\hline
    Generated  & & 19360 & 44827100 & & 26307 & 20855900 \\ \hline
    Missing mass  (GeV)& $80<M_{miss}<140$ & 15466 & 6214050 & $50<M_{miss}<240$ & 23202 & 5627040 \\ \hline
    Transverse $P$ (GeV)& $20<P_{T}<70$  & 13727 & 549340 & $10<P_{T}<140$ & 22648 & 2271090 \\ \hline
    Longitudinal $P$ (GeV)& $|P_{L}|<60$ & 13342 & 392401 & $|P_{L}|<130$ & 22459 & 2051010 \\ \hline
    \# of charged tracks  & $N_{chd}>10$ & 12936 & 374877 & $N_{chd}>10$  & 21270 & 1936220 \\ \hline
    Maximum $P$ (GeV)& $P_{max}<30$ & 11743 & 205038 & $P_{max}<60$ & 20556 & 1167050 \\ \hline
    $Y_{23}$ value & $Y_{23}<0.02$ & 7775 & 74439 & $Y_{23}<0.02$ & 14992 & 465461 \\ \hline
    $Y_{12}$ value & $0.2<Y_{12}<0.8$ & 7438 & 62584 & $0.2<Y_{12}<0.8$ & 14500 & 413762 \\ \hline
    Di-jet mass (GeV)& $100<M_{jj}<130$ & 6691 & 19061 & $100<M_{jj}<130$ & 12334 & 71918 \\ \hline
    Likelihood ratio & $LR>0.165$ & 6293 & 10940 & $LR>0.395$ & 9543 & 11092 \\
    \hline\hline
    Significance (Efficiency) & $S/\sqrt{S+B}$ & \multicolumn{2}{c|}{47.9 (32.5\%)} &$S/\sqrt{S+B}$ & \multicolumn{2}{c|}{66.4 (36.3\%)}\\
    \hline
   \end{tabular}
  \end{center}
 \end{table}
  
 \subsection{Hadronic channel ($q\bar{q}H$)}

 For hadronic channel analysis,
 particles in the event are first forcibly clustered into four jets.
 Next, a Higgs and $Z$ candidate dijet pair that minimize the
 following $\chi^{2}$ formula are selected from the four jets:
 \begin{equation}
  \chi^{2} = \left(\frac{M_{j_{1}j_{2}}-M_{Z}}{\sigma_{Z}}\right)^2+\left( \frac{M_{j_{3}j_{4}}-M_{H}}{\sigma_{H}} \right)^2,
 \end{equation}
 where $M_{j_{1}j_{2}/j_{3}j_{4}}$, $M_{Z/H}$ represent the dijet invariant masses paired from the four jets ($j_{1-4}$)
 and the $Z$ and Higgs masses, respectively.
 Here $\sigma_{Z}=4.7~{\rm GeV}$ and $\sigma_{H}=4.4~{\rm GeV}$ are used for $\sqrt{s}=250~{\rm GeV}$ and
 $\sigma_{Z/H}=3.8~{\rm GeV}$ for $\sqrt{s}=350~{\rm GeV}$.
 They are determined from the dijet mass distribution reconstructed from the true MC information.
 After the jet pairing, background reductions are applied.

 To select the four-jet-like events,
 cuts on the number of charged tracks $N_{charged}$ and jet clustering parameter $Y_{34}$ are applied.
 $Y_{34}$ is the minimum scaled jet mass $y$ required for four-jet clustering.
 The leptonic backgrounds ($\llll$, $\llqq$) are reduced effectively by these selections.
 In addition, cuts on the thrust and thrust angle are applied to reduce the $ZZ$ background,
 utilizing the difference between the event shape of the signal (spherical) and $ZZ$, $\qq$ (back-to-back).
 The numbers of $\qqqq$ and $q\bar{q}$ background events are reduced by a cut on the angle
 between the Higgs candidate jets ($\theta_{H}$).
 The $WW$ and $ZZ$ backgrounds are further suppressed by cuts on the Higgs and $Z$ candidates
 after the kinematical constraint fit is applied to the four-jet system as follows.
 Each jet is parameterized by $E_{j_{i}}$, $\theta_{i}$, and $\phi_{i}$ ($i=1-4$)
 and fitted with constraints on the total energy ($\sum_{i} E_{j_i} =\sqrt{s}$),
 the total momentum ($\sum_{i} \vec{P}_{j_{i}} = 0$),
 and Higgs and $Z$ mass difference ($|M_{j_{1}j_{2}}-M_{j_{3}j_{4}}|=|M_{H}-M_{Z}|$),
 where $E_{j_{i}}$, $P_{j_{i}}$, $\theta_{i}$, and $\phi_{i}$ are
 the energy, momentum, and theta and phi angles of the $i$-th jet, respectively.
 After these cuts are applied,
 an additional cut is applied on the $LR$ derived from the following input variables:
 thrust, $\cos{\theta_{\rm thrust}}$, minimum angle between all jets ($\theta_{min}$),
 number of particles in Higgs candidate jets, fitted $Z$ mass, and fitted Higgs mass.
 The likelihood cut position is selected to maximize signal significance; $LR > 0.375$ for 250 GeV
 and $LR > 0.15$ for 350 GeV.
 All background reduction procedures are summarized in Table~\ref{table:qqh}.
 The background fractions after all cuts are 80\% $\qqqq$ and 20\% $q\bar{q}$ at 250 GeV
 and 60\% $\qqqq$, 30\% $q\bar{q}$ and 10\% $t\bar{t}$ at 350 GeV.
 \begin{table}[htbp]
  \begin{center}
   \caption{Summary of $q\bar{q}H$ channel background reduction assuming ${\cal L}=250~{\rm fb^{-1}}$ with $P(e^{-}, e^{+})=(-0.8, +0.3)$.}
   \label{table:qqh}
   \begin{tabular}{|c|c|r|r|c|r|r|}
    \hline
    CM energy (GeV) &\multicolumn{3}{c|}{250} & \multicolumn{3}{c|}{350} \\    
    \hline
    Cut names & condition & Sig.& Bkg. & condition & Sig. & Bkg.\\
    \hline\hline
    Generated     & & 52507 & 45904900 & & 36099 & 22210900 \\ \hline
    $\chi^{2}$    & $\chi^{2}<10$ & 32447 & 2608980 & $\chi^{2}<10$ & 20207 & 1034810 \\ \hline
    \# of charged tracks & $N_{chd}>4$ & 25281 & 1120950 & $N_{chd}>4$ & 14900 & 305649 \\ \hline
    $Y_{34}$ value & $-\log(Y_{34})>2.7$ & 25065 & 1002125 & $-\log(Y_{34})>2.7$ & 14543 & 250995 \\ \hline
    thrust       & $\rm thrust<0.9 $ & 24688 & 935950 & $\rm thrust<0.85$ & 13522 & 144560 \\ \hline
    thrust angle & $|\cos\theta_{\rm thrust}|<0.9$ & 21892 & 696201 & $|\cos\theta_{\rm thrust}|<0.9$ & 12523 & 107025 \\ \hline
    Higgs jets angle & $105^{o}<\theta_{H}<160^{o}$ & 20062 & 622143 & $70^{o}<\theta_{H}<120^{o}$ & 11185 & 77659 \\ \hline
    $Z$ di-jet mass (GeV)& $80<M_{Z}<100$ & 16359 & 411863 & $80<M_{Z}<100$ & 9468 & 45671 \\ \hline
    $H$ di-jet mass (GeV)& $105<M_{H}<130$ & 16359 & 411863 & $105<M_{H}<130$ & 9451 & 44399 \\ \hline
    Likelihood ratio & $LR>0.375$ & 13726 & 166807 & $LR>0.15$ & 8686 & 25393\\
    \hline\hline
    Significance (Efficiency) & $S/\sqrt{S+B}$ & \multicolumn{2}{c|}{32.3 (26.1\%)} &$S/\sqrt{S+B}$ & \multicolumn{2}{c|}{47.1 (24.1\%)}\\
    \hline
   \end{tabular}
  \end{center}
 \end{table}
 
 \subsection{Leptonic channel ($\ell^{+}\ell^{-}H$)}

 For leptonic channel analysis,
 we considered the cases where the lepton is an electron or a muon.
 We considered only the $\ell\ell qq$ and $\ell \nu qq$ background processes.
 First, the following cuts were applied to selected isolated leptons:

  \begin{itemize}
   \item{Lepton isolation: $E_{cone} < 20~{\rm GeV}$ (cone angle: $10^{\circ}$)},
   \item{Lepton track momentum: }\\
	$10 < E_{lep} < 90 ~{\rm GeV}$ at $\sqrt{s}=250~{\rm GeV}$,\\
	$10<E_{lep}<160~{\rm GeV}$ at $\sqrt{s}=350~{\rm GeV}$,
  \end{itemize}
 where $E_{cone}$ is the energy sum for particles within $10^{o}$ of the lepton.
 The prompt lepton has a smaller $E_{cone}$ than nonprompt leptons.
 Electrons and muons are identified from their charged tracks as follows:
  \begin{itemize}
   \item{Electron ID: $\displaystyle\frac{E_{ECAL}}{E_{Total}}>0.9$, $0.7 < \displaystyle\frac{E_{Total}}{P}<1.2$ }
   \item{Muon ID: $\displaystyle\frac{E_{ECAL}}{E_{Total}}<0.5$, $\displaystyle\frac{E_{Total}}{P}<0.4$},
  \end{itemize}
 where $E_{ECAL}$, $E_{Total}$ and $P$ denote the $ECAL$ energy associated with a track,
 total energy deposited in the ECAL and HCAL, and track momentum, respectively.
 If there are more than two isolated lepton candidates after the electron or muon identification,
 a pair whose invariant mass is closest to $Z$ is selected.
 After  dilepton identification,
 forced two-jets clustering is applied to the remaining particles and the following selections are applied.
 First, a dilepton mass ($M_{\ell\ell}$) cut, which should be consistent with the $Z$ mass, is applied:
 $70 < M_{\ell\ell} < 110~{\rm GeV}$ for electrons and $70 < M_{\ell\ell} < 100~{\rm GeV}$ for muons.
 Because the $ZZ$ or $WW$ backgrounds are boosted to the forward region compared to the signal,
 a cut on the $Z$ direction is applied: $|\cos{\theta_{Z}}| < 0.8$.
 Finally, cuts on dijet mass ($M_{jj}$) and a mass recoil to the lepton pair ($M_{rec}$) are applied to select the Higgs signal:
 $100 < M_{jj} < 140~{\rm GeV}$ and $70 < M_{rec} < 140~{\rm GeV}$ for electrons;
 $115 < M_{jj} < 140~{\rm GeV}$ and $70 < M_{rec} < 140~{\rm GeV}$ for muons.
 The background reduction procedures for the leptonic channel are summarized in Table~\ref{table:llh}.
 After all cuts were applied, the background was dominated by the $\llqq$ whereas the $\nlqq$ was well suppressed.
 
 \begin{table}[htbp]
  \begin{center}
   \caption{Summary of background reduction in the $eeH$ and $\mu\mu H$ channels  assuming ${\cal L}=250~{\rm fb^{-1}}$ with $P(e^{-}, e^{+}=(-0.8, +0.3))$.}
   \label{table:llh}
   \begin{tabular}{|c|c|c|r|r|c|r|r|}
    \hline
    CM energy (GeV) & &\multicolumn{3}{c|}{250} & \multicolumn{3}{c|}{350} \\
    \hline
    Cut names & $e/\mu$ & condition & Sig.& Bkg. & condition & Sig. & Bkg.\\
    \hline\hline
    Generated & $e$ & & 3137 & 4512520 & & 2740& 3822410 \\
              & $\mu$ & & 2917 & 4512520 & & 1789& 3822410\\ \hline
    \# of $e/\mu$ track ID & $e$ & $N_{e}>=2$ & 2717 & 204403 & $n_{e}>=2$ & 2270& 179580\\
                           & $\mu$ & $N_{\mu}>=2$ & 2668 & 28175 & $N_{\mu}>=2$ & 1631& 23598\\ \hline
    Di-lepton mass (GeV) & $e$ & $70<M_{\ell\ell}<110$ & 2208 & 34162 & $70<M_{\ell\ell}<110$ & 1425& 51436\\
                         & $\mu$ & $80<M_{\ell\ell}<110$ & 2287 & 12901 & $80<M_{\ell\ell}<100$ & 1406& 13313\\    \hline
    $Z$ direction & $e$ & $|\cos{\theta}|<0.8$& 1797 & 21600 & $|\cos{\theta}|<0.8$& 1192& 20874\\
                        & $\mu$ & $|\cos{\theta}|<0.8$& 1889& 8036& $|\cos{\theta}|<0.8$& 1203& 6250\\ \hline
    Di-jet mass (GeV) & $e$ & $100<M_{jj}<140$& 1394 & 2721 & $110<M_{jj}<140$ & 865& 2019\\
                      & $\mu$ & $115<M_{jj}<140$& 1445 & 1955 & $115<M_{jj}<140$ & 855& 1197\\ \hline
    Recoil mass (GeV) & $e$ & $70<M_{rec}<140$& 1184 & 1607 & $70<M_{rec}<140$ & 567& 590\\
                      & $\mu$ & $70<M_{rec}<140$& 1365 & 983 & $70<M_{rec}<140$& 638&465 \\ \hline
    \hline\hline
    Significance (Efficiency) & $e$   & $S/\sqrt{S+B}$ & \multicolumn{2}{c|}{ 22.4 (37.8\%)} &$S/\sqrt{S+B}$ & \multicolumn{2}{c|}{ 16.7 (20.7\%)}\\
                              & $\mu$ &  & \multicolumn{2}{c|}{ 28.2 (46.8\%)} & & \multicolumn{2}{c|}{ 19.2 (35.7\%)}\\    
    \hline
   \end{tabular}
  \end{center}
 \end{table}
 
 \section{Branching ratio measurement} \label{sec:template}

 After event selection,
 the measurement accuracies of the Higgs BRs to $b\bar{b}$, $c\bar{c}$, and $gg$ are evaluated on the basis of a template fitting to the flavor likeness of the Higgs dijets obtained by using the LCFIVertexing package~\cite{LCFI}.
 The probabilities of $b$ and $c$ quarks for each jet [$b_{i},~c_{i}~(i=1, 2)$] are calculated
 in LCFIVertex using neural net training with a $Z \to q\bar{q}$ samples at the $Z$-pole.
 In addition, another $c$ probability ($bc_{1, 2}$) is also calculated
 whose neural-net is trained only with $Z\to b\bar{b}$ sample as the background.
 For Higgs dijets, we define the flavor likeness $X$ ($X=b,~c,~bc$) as follows from the $x_{i}$ [$x_{i}=b_{i},~c_{i},~bc_{i}~(i=1,2)$] flavor probability of each jet:
 \begin{equation} \label{eq:flavor-likeness}
  X = \frac{x_{1}x_{2}}{x_{1}x_{2}+(1-x_{1})(1-x_{2})}.
 \end{equation}
 
\begin{figure}[htbp]
  \begin{center}
   \includegraphics[width=0.7\textwidth]{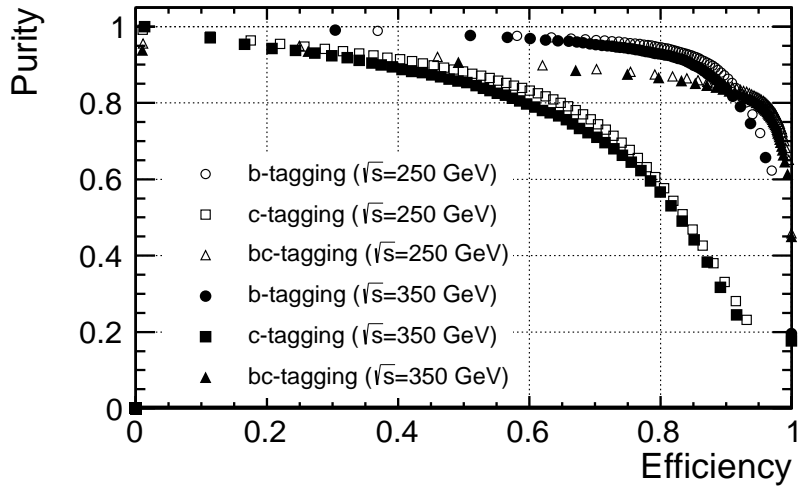}
  \end{center}
  \caption{Flavor tagging performance at CM energies of 250 and 350 GeV in $ZZ \to \nu\bar{\nu}q\bar{q}$ sample. Horizontal axis shows the efficiency of $b/c$ jets; vertical axis shows the purity of tagged $b/c$ jets.}
  \label{fig:flavor-tagging}
 \end{figure}

 The flavor tagging performance in the $ZZ \to \nu\bar{\nu}q\bar{q}$ sample
 at the $\sqrt{s} = 250~{\rm and}~350~{\rm GeV}$ is shown in Fig.~\ref{fig:flavor-tagging}.
 The $ZZ\to \nu\bar{\nu}q\bar{q}$ samples are compared for each CM energy
 because they form the same final state as $Z\to q\bar{q}$,
 which was used to train the flavor tagging neural network.
 Figure~\ref{fig:flavor-tagging} shows that no significant
 difference in the flavor tagging performance at  $\sqrt{s}$ = 250 and 350~GeV
 is observed for any of the flavors.\\

\begin{figure}[htbp]
  \begin{center}
   \includegraphics[width=\textwidth]{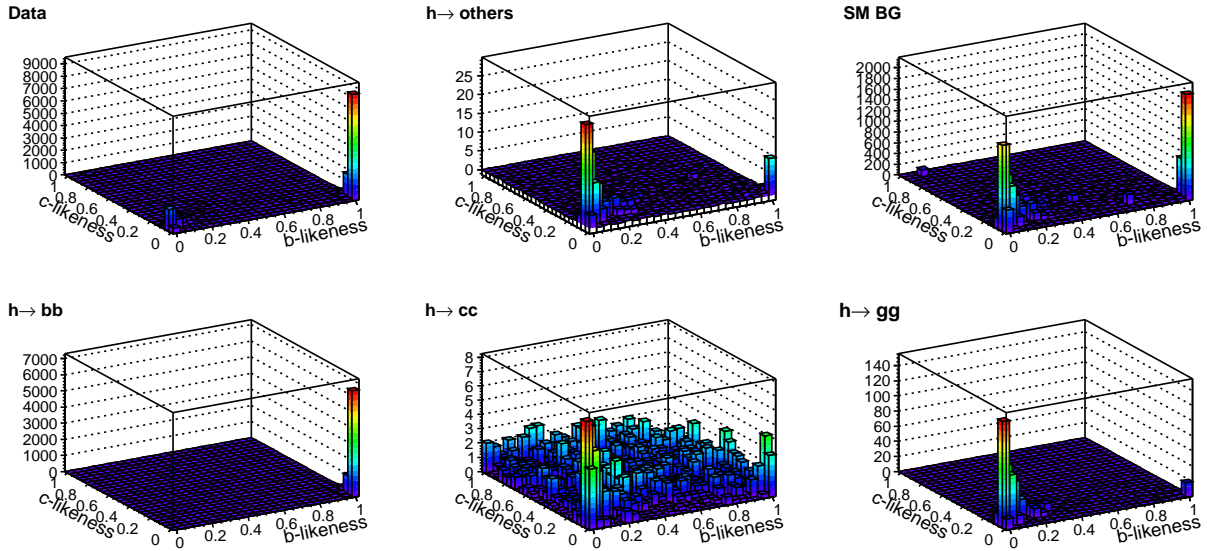} 
   \caption{Two-dimensional images of three-dimensional template samples for $b$-likeness vs. $c$-likeness}
   \label{fig:2Dsample}
  \end{center}
 \end{figure}

 To evaluate the measurement accuracy of the BRs,
 the $b$-, $c$-, and $bc$-likenesses of the selected events were binned in a three-dimensional histogram and fitted with those of the template samples,
 which consist of $H\to b\bar{b}$, $c\bar{c}$, and $gg$ and other background processes.
 Figure~\ref{fig:2Dsample} shows the three-dimensional histogram projected to the two-dimensional $b$- and $c$-likeness axes for the hadronic channel.
 The probability of entries in each template sample bin is expected to be given by the Poisson statistics:
 \begin{equation}\label{eq:prob}
  P_{ijk} = \frac{\mu^{n}e^{-\mu}}{n!}~\left(n\equiv N^{data}_{ijk},~\mu\equiv N^{template}_{ijk}\right),
 \end{equation}
 where $P_{ijk}$ and $N^{data}_{ijk}$ are the probability of entries
 and the number of data entries at the $(i, j, k)$ bin, respectively.
 $N^{template}_{ijk}$ is given by
 \begin{equation}
  N^{template}_{ijk} = \sum_{s=b\bar{b},c\bar{c},gg} r_{s}\cdot N^{s}_{ijk} + N_{ijk}^{bkg},
 \end{equation}
 where $N^{s}_{ijk}$ is the number of entries at the $(i,j,k)$ bin in each $H\to b\bar{b}$, $c\bar{c}$, and $gg$ template;
 $N^{bkg}_{ijk}$ is the number of entries in the background template sample, which is the sum of the SM background events and the Higgs-to-nonhadronic decay events. Furthermore, 
 $r_{b\bar{b}}$, $r_{c\bar{c}}$, and $r_{gg}$ are the parameters to be determined by the template fitting.
 They are defined as the Higgs branching fractions to $H\to b\bar{b}$, $c\bar{c}$ and $gg$, respectively, normalized by that of the SM,
 \begin{equation}\label{eq:r_s}
  r_{s} = \frac{\sigma \cdot Br\left(H\to s \right)}{\sigma^{SM} \cdot Br\left(H\to s \right)^{SM}}~(s=b\bar{b},~c\bar{c},~gg).
 \end{equation}
 Here $\sigma$ is the Higgs production cross section and
 $\sigma^{SM}$ and $Br(H\to s)^{SM}$ are the cross section and branching fraction in the SM, respectively.
 From Eq.~(\ref{eq:r_s}), the measurement accuracies of $\sigma \cdot Br$ are obtained as follows;
 \[
  \frac{\Delta \left(\sigma\cdot Br\right)}{\sigma \cdot Br}(H\to s) = \frac{\Delta r_{s}}{r_{s}}~(s = b\bar{b},~c\bar{c},~gg).
 \]
 The $r_{s}$'s values were determined by a binned log likelihood fitting, where each bin probability is given by Eq.~(\ref{eq:prob}).

\begin{figure}[htbp]
  \begin{center}
   \includegraphics[width=\textwidth]{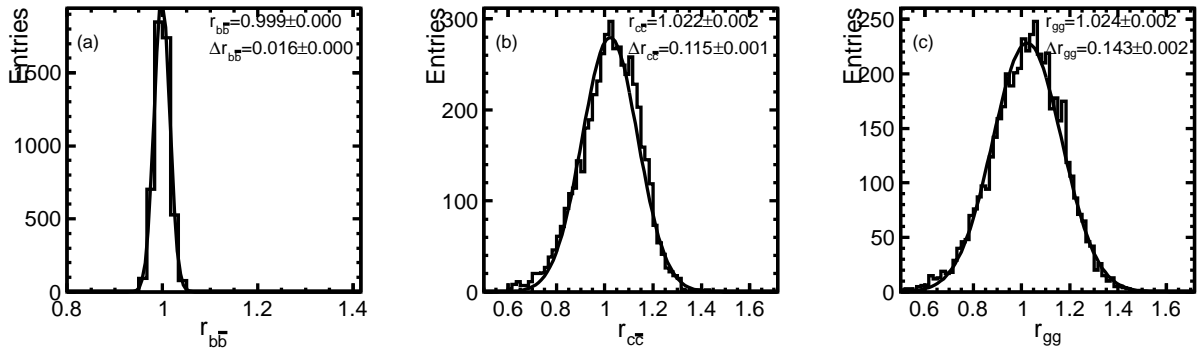} 
   \caption{Typical examples of (a) $r_{b\bar{b}}$, (b) $r_{c\bar{c}}$, and (c) $r_{gg}$ distributions.}
   \label{fig:rxx}
  \end{center}
 \end{figure}

 On the basis of the three-dimensional (3D) histogram, 5000 toy MC events were generated using the Poisson distribution function for each bin,
 which were fitted to obtain $r_{b\bar{b}}$, $r_{c\bar{c}}$, and $r_{gg}$.
 The number of bins in the 3D histogram were optimized to minimize the statistical fluctuation in the fitted results caused by low-statistic bins.
 Bins with fewer than one entry were not used for the fitting.
 The distributions of $r_{b\bar{b}}$, $r_{c\bar{c}}$, and $r_{gg}$ for template fitting to
 1000 toy MC events are shown in Fig.~\ref{fig:rxx}.
 The error in $r_{s}$ is determined by the Gaussian fittings to these distributions,
 which are shown in Tables~\ref{table:template_250} and \ref{table:template_350} for CM energies of 250 and 350 GeV, respectively.
 
 The tables also show the accuracies after correction of the total cross section.
 From a study of the recoil mass in the process of $e^{+}e^{-}\to eeH$ and $\mu\mu H$,
 the accuracy of the total cross section ($\Delta \sigma /\sigma $) was estimated to be 2.5\% at 250~GeV~\cite{ILD, ZH_recoil}.
 For 350 GeV, we assumed an accuracy of 3.5\% because the recoil mass measurement relies on the $ZH$ process,
 whose cross section is inversely proportional to the square of the CM energy;
 thus, the accuracy of the total cross section measurement would be inversely proportional to the CM energy.

 From Tables~\ref{table:template_250} and \ref{table:template_350},
 we see that the Higgs cross section times branching ratio
 can be measured at about 1\% for $H\to b\bar{b}$ and 7 to 9\% for $H \to c\bar{c}$ and $gg$.
 The measurement is approximately $10-20\%$ better at 350 GeV than at 250 GeV.
 The instantaneous luminosity at 350 GeV is 25\%
 greater than that at 250 GeV according to the ILC beam parameters.
 Thus, for an equal running time, measurements at 350 GeV will give us
 about $20-30\%$ better accuracy than those at 250 GeV.
 On the other hand,
 the accuracy of the BR to $b\bar{b}$, $\Delta Br/Br(H\to b\bar{b})$, is limited
 by the total cross section ambiguity; thus, measurement at 250 GeV gives us better results than that at 350 GeV.
 In the other decay channels,
 comparable BR measurements are possible even if the same integrated luminosities are assumed.
 
\begin{table}[htbp]
 \begin{center}
  \caption{Summary of template fitting results $r_{s}$ and accuracies of ($\sigma\cdot Br$) and $Br$ after correcting $\sigma$ for an accuracy of 2.5\% at $\sqrt{s}=250~{\rm GeV}$
  assuming ${\cal L}=250~{\rm fb^{-1}}$ with $(e^{-}, e^{+})=(-0.8, +0.3)$.}
\label{table:template_250}
  \begin{tabular}{c|c|c|c|c|c}
   \hline
   & $\nu\bar{\nu}H$ & $q\bar{q}H$ & $e^{+}e^{-}H$ &$\mu^{+}\mu^{-}H$ & comb.\\
   \hline\hline
   $r_{b\bar{b}}$ & 1.00$\pm$0.02 & 1.00$\pm$0.01 & 1.00$\pm$0.04 & 1.00$\pm$0.03 & 1.00$\pm$0.01\\
   \hline
   $r_{c\bar{c}}$ & 1.02$\pm$0.11 & 1.01$\pm$0.10 & 1.02$\pm$0.27 & 1.01$\pm$0.23 & 1.02$\pm$0.07 \\
   \hline
   $r_{gg}$       & 1.02$\pm$0.14 & 1.02$\pm$0.13 & 1.05$\pm$0.33 & 1.02$\pm$0.24 & 1.02$\pm$0.09 \\
   \hline
   $\frac{\Delta (\sigma\cdot Br)}{\sigma \cdot Br}(H\to b\bar{b})$ (\%) &  1.7 &  1.5 &  3.8 &  3.3 & 1.0\\
   \hline				      
   $\frac{\Delta (\sigma\cdot Br)}{\sigma \cdot Br}(H\to c\bar{c})$ (\%) & 11.2 & 10.2 & 26.8 & 22.6 & 6.9\\ 
   \hline				      
   $\frac{\Delta (\sigma\cdot Br)}{\sigma \cdot Br}(H\to gg)$ (\%)       & 13.9 & 13.1 & 31.3 & 33.0 & 8.5\\ 
   \hline
   $\frac{\Delta Br}{Br}(H\to b\bar{b})$ (\%) &  3.0 &  2.9 &  5.7 &  4.5 & 2.7\\
   \hline				      		      	            
   $\frac{\Delta Br}{Br}(H\to c\bar{c})$ (\%) & 11.4 & 10.5 & 31.3 & 22.8 & 7.3\\
   \hline				      		      	            
   $\frac{\Delta Br}{Br}(H\to gg)$ (\%)       & 14.2 & 13.3 & 33.1 & 24.0 & 8.9\\  
   \hline
  \end{tabular}
 \end{center}
\end{table}

\begin{table}[htbp]
 \begin{center}
  \caption{Summary of template fitting results $r_{s}$ and accuracies of ($\sigma\cdot Br$) and $Br$ after correcting $\sigma$ for an accuracy of 3.5\% at $\sqrt{s}=350~{\rm GeV}$ assuming ${\cal L}=250~{\rm fb^{-1}}$ with $(e^{-}, e^{+})=(-0.8, +0.3)$.}
  \label{table:template_350}
  \begin{tabular}{c|c|c|c|c|c}
   \hline
   & $\nu\bar{\nu} H$ & $q\bar{q}H$ & $e^{+}e^{-}H$ &$\mu^{+}\mu^{-}H$ & comb.\\
   \hline\hline
   $r_{b\bar{b}}$ & 1.00$\pm$0.01 & 1.00$\pm$0.02 & 1.00$\pm$0.05 & 1.00$\pm$0.05 & 1.00$\pm$0.01 \\
   \hline					       
   $r_{c\bar{c}}$ & 1.02$\pm$0.11 & 1.01$\pm$0.10 & 1.02$\pm$0.31 & 1.04$\pm$0.32 & 1.01$\pm$0.06 \\
   \hline					       
   $r_{gg}$       & 1.02$\pm$0.14 & 1.04$\pm$0.14 & 1.04$\pm$0.37 & 1.03$\pm$0.34 & 1.02$\pm$0.07 \\
   \hline
   $\frac{\Delta (\sigma\cdot Br)}{\sigma \cdot Br}(H\to b\bar{b})$ (\%)  & 1.4 &  1.5 &  5.3 &  5.1 & 1.0\\
   \hline					   
   $\frac{\Delta (\sigma\cdot Br)}{\sigma \cdot Br}(H\to c\bar{c})$ (\%)  & 8.6 & 10.1 & 30.5 & 30.9 & 6.2\\
   \hline					 
   $\frac{\Delta (\sigma\cdot Br)}{\sigma \cdot Br}(H\to gg)$ (\%)        & 9.2 & 13.7 & 35.8 & 33.0 & 7.3\\
   \hline
   $\frac{\Delta Br}{Br}(H\to b\bar{b})$ (\%)  & 3.8 &  3.8 &  6.4 &  6.2 & 3.6\\
   \hline					 
   $\frac{\Delta Br}{Br}(H\to c\bar{c})$ (\%)  & 9.2 & 10.6 & 30.7 & 31.1 & 7.2\\
   \hline					 
   $\frac{\Delta Br}{Br}(H\to gg)$ (\%)        & 9.8 & 14.1 & 36.0 & 33.2 & 8.1\\
   \hline 
  \end{tabular}
 \end{center}
\end{table}
 
 \section{Conclusion}

 The measurement accuracy of the Higgs branching fractions, $H\to b\bar{b}$, $c\bar{c}$, and $gg$, were evaluated at $\sqrt{s}=250~{\rm GeV}$ and $350~{\rm GeV}$.
 In terms of signal significance, $\sqrt{s}=350~{\rm GeV}$ yields better background suppression than $\sqrt{s}=250~{\rm GeV}$ for each channel.
 The combined results for measurement accuracies of the Higgs cross section times BRs
 ($\Delta (\sigma \cdot Br)/\sigma\cdot Br$) to $H\to b\bar{b}$, $c\bar{c}$, and $gg$
 are 1.0\%, 6.9\%, and 8.5\% at CM energies of 250 GeV
 and 1.0\%, 6.2\%, and 7.3\% at 350 GeV, respectively, 
 assuming the same integrated luminosity of $250~{\rm fb}^{-1}$.
 At the ILC, the total Higgs cross-section $\sigma$ is measured using the $Z$ recoil mass process.
 Using $\Delta \sigma/\sigma=2.5\%$ for 250 GeV and assuming it is 3.5\% at 350 GeV, 
 Higgs BRs ($\Delta Br/Br$) to $b\bar{b}$, $c\bar{c}$, and $gg$
 are derived as
 2.7\%, 7.3\%, and 8.9\% at CM energies of 250 GeV and
 as 3.6\%, 7.2\%, and 8.1\% at 350 GeV.
 Therefore, we conclude that
 the Higgs cross section times BR ($Br\times \sigma$) can be measured better at 350 GeV than at 250 GeV
 owing to the higher S/N at the higher energy.
 However, when the accuracy of the total cross section measurement
 by recoil mass measurement is considered,
 BR of $H\to b\bar{b}$ can be measured better at 250 GeV,
 even if the integrated luminosity is the same at both energies.
  
 \section*{Acknowledgment}
 The authors thank the members of the ILC physics subgroup
 for useful discussions of this work and those of the ILD software and optimization group,
 who maintain the software and MC samples used in this work.
 This work was supported in part by Creative Scientific Research Grant No. 18GS0202 from the Japan Society for Promotion of Science (JSPS), the JSPS Core University Program, and JSPS Grant-in-Aid for Scientific Research No. 22244031.

\end{document}